\documentstyle[amsmath,amssymb,aps,psfig,multicol]{revtex}
\begin{document}
\setlength{\topmargin}{-1.3cm}
\draft
\title{Construction and properties of assortative random networks}
\author{R. Xulvi-Brunet and I.M. Sokolov}
\address{Institut f\"{u}r Physik, Humboldt Universit\"{a}t zu Berlin, 
         Newtonstra\ss e 15, D-12489 Berlin, Germany}
\maketitle

\bigskip

\begin{abstract}
Many social networks exhibit assortative mixing so that the predictions of 
uncorrelated models might be inadequate. To analyze the role of assortativity 
we introduce an algorithm which changes correlations in a network and produces 
assortative mixing to a desired degree. This degree is governed by one 
parameter $p$. Changing this parameter one can construct networks ranging from 
fully random ($p = 0$) to totally assortative ($p = 1$). We apply the 
algorithm to a Barab\'asi-Albert scale-free network and show that the degree 
of assortativity is an important parameter governing geometrical and transport 
properties of networks. Thus, the diameter of the network and the clustering 
coefficient increase dramatically with the degree of assortativity. Moreover, 
the concentration dependences of the size of the giant component in the node 
percolation problem for uncorrelated and assortative networks are strongly 
different.
\end{abstract}



\begin{multicols}{2}

\section*{Introduction}

Complex networks have recently attracted a burst of interest as an 
indispensable tool of description of different complex systems. Thus, 
technological webs as the Internet and the World Wide Web, as well as other 
natural and social systems like intricate chemical reactions in the 
living cell, the networks of scientific and movie actors' collaborations, and 
even human sexual contacts have been successfully described through scale-free
networks, networks with the degree distribution $P(k) \sim k^{-\gamma}$
\cite{baretalb,dorogov}. The degree distribution $P(k)$ is one of the 
essential measures used to capture the structure of a network, and gives the 
probability that a node chosen at random is connected with exactly $k$ other 
vertices of the network. 

Recently, it was pointed out that the existence of degree correlations among 
nodes is an important property of the real networks
\cite{NE,Newm,Alex,Cap,Pastor,Park,Berg,Goh,Mas,Kra,Clus,Call,Juyong}. 
Thus, many social networks show that nodes having many connections tend to be 
connect with other highly connected nodes \cite{Newm,Cap}. In the literature 
this characteristics is usually dented as assortativity, or assortative mixing.
On the other hand, technological and biological networks often have the 
property that nodes with high degree are preferably connected with ones with 
low degree, a property referred to as dissortativity \cite{NE,Pastor}. Such 
correlations have an important influence on the topology of networks, and 
therefore they are essential for the description of spreading phenomena, like 
spreading of information or infections, as well as for the robustness of 
networks against intentional attack or random breakdown of their elements 
\cite{Bog,Egui,Bo,Schwartz,Vaz,Mor}. 

In order to assess the role of correlations, especially of the assortative 
mixing, several studies have proposed procedures to build correlated 
networks \cite{NE,Rame,Jo,Mar}. The most general of them are the ones proposed
by Newman \cite{NE}, and by Bogu\~n\'a and Pastor-Satorras \cite{Mar}, who 
suggest two different ways to construct general correlated networks with 
prescribed correlations. Following the same goal, we however adopt a different
perspective in this paper. We propose a simple algorithm producing assortative
mixing, in which, instead of putting correlations by hand, we only try impose 
the intuitive condition that ``nodes with similar degree connects preferably''.
We then investigate the correlations which come out of our simple model. Thus,
we present an algorithm, governed by the only parameter $p$, capable to 
generate assortative correlations to a desired degree. In order to study the 
effect of the assortative mixing, we apply our algorithm to a 
Barab\'asi-Albert scale-free network \cite{BA-model}, the one leading to the 
degree distribution $P(k) \sim k^{-3}$, and investigate the properties of 
the emerging networks in some detail.

\section*{The algorithm}

In what follows we treat undirected networks. Starting from a given network,
at each step two links of the network are chosen at random, so that the four 
nodes, in general, with different degrees, connected through the links two by 
two are considered. The step of our algorithm looks as follows. The four nodes
are ordered with respect to their degrees. Then, with probability $p$, the 
links are rewired in a such a way that one link connects the two nodes with 
the smaller degrees and the other connects the two nodes with the larger 
degrees, otherwise the links are randomly rewired (Maslov-Sneppen algorithm 
\cite{Mas}). In the case when one, or both, of these new links already existed
in the network, the step is discarded and a new pair of edges is selected. This
restriction prevents the appearance of multiple edges connecting the same pair
of nodes. A repeated application of the rewiring step leads to an assortative 
version of the original network. Note that the algorithm does not change the 
degree of the nodes involved and thus the overall degree distribution in the 
network. Changing the parameter $p$, it is possible to construct networks with 
different degree of assortativity.

\section*{Correlations and assortativity}

Let ${\cal E}_{ij}$ be the probability that a randomly selected edge of the 
network connects two nodes, one with degree $i$ and another with degree $j$. 
The probabilities ${\cal E}_{ij}$ determine the correlations of the network.
We say that a network is uncorrelated when 
\begin{equation}
  {\cal E}_{ij} = (2-\delta_{ij})\frac{iP(i)}{\langle i \rangle}\frac{jP(j)}
           {\langle j \rangle} := {\cal E}^r_{ij} \ \ \ ,    \label{primera}
\end{equation}
i.e, when the probability that a link is connected to a node with a certain 
degree is independent from the degree of the attached node. 
Here $ \langle i \rangle = \langle j \rangle $ denotes the
first moment of the degree distribution. 

Assortativity means that highly connected nodes tend to be connected to each
other with a higher probability than in an uncorrelated network. Moreover, the 
nodes with similar degrees tend to be connected with larger probability than 
in the uncorrelated case, i. e., 
${\cal E}_{ii} > {\cal E}^r_{ii}$ \ $\forall i$. The 
degree of assortativity of a network can thus be characterized by the quantity
\cite{NE}:
\begin{equation} 
   {\cal A} = \frac{\sum_{i} {\cal E}_{ii} - \sum_{i} {\cal E}^r_{ii}}
                {1-\sum_{i} {\cal E}^r_{ii}}  \ \ \ ,  \label{segona}
\end{equation}
which takes the value $0$ when the network is uncorrelated and the
value $1$ when the network is totally assortative. (Note that finite-size 
effects and the constraint that no vertices are connected by more than one 
edge bound ${\cal A}$ from above by the values lower than 1 \cite{Snep}).

Now, starting from the algorithm generator, we can obtain a theoretical 
expression for ${\cal E}_{ij}$ as a function of $p$. Let $E_{ij}$ be the number
of links in the network connecting two nodes, one with degree $i$ and another 
with degree $j$, so that ${\cal E}_{ij}= E_{ij}/L$, where $L$ is the total 
number of links of the network. (Since undirected networks satisfy 
$E_{ij}=E_{ji}$, the restriction $i \le j$ can be imposed without loss of 
generality). We now define the variable
\begin{equation}
   F_{ln} = \sum^n_{r=l} \sum^n_{s=r} E_{rs} \ \ \ \ \ 
                       r \le s \ \ ; \ \ l \le n \ \ \ .  \label{tercera}
\end{equation}
A careful analysis of the algorithm reveals that, every time the rewiring 
process is applied, $F_{ln}$ either does not change, or changes increasing or 
decreasing by unity. We can then calculated the probabilities that it changes,
i. e., that $F_{ln} \rightarrow F_{ln}+1$ or $F_{ln} \rightarrow F_{ln}-1$. 
Here, the effect of multiple edges can be disregarded since they are rare in 
the thermodynamical limit of infinite networks. Taking into account all 
corresponding possibilities, we obtain for the probabilities of changes the 
following expressions:
\[
  \left( X_{ln}-f_{ln} \right)^2 + p \left( X_{ln}-f_{1n}+f_{1,l-1} \right)^2  
\]
for $F_{ln} \rightarrow F_{ln}+1$ and
\[
  f_{ln} \left[ (1-p)(1-2X_{ln})+p(X_{1,l-1}-f_{1,l-1}-f_{1n})+f_{ln} \right]
\]
for $F_{ln} \rightarrow F_{ln}-1$.
Here $f_{ln}=F_{ln}/L$, and $X_{ln}$ is given by:
\[
    X_{ln} = \frac{1}{\langle k \rangle}\sum^n_{k=l} k P(k) \ \ \ \ \ \ \ 
              \ \ \ \ \ l \le n \ \ \ .  
\]
(Note that $X_{ln}$ and $f_{ln}$ vanish when one of the indices is smaller 
than $1$, the minimal tolerated degree). Using this, we can calculate the 
expected value for $f_{ln}$. The process of repeated applying our algorithm 
corresponds to an ergodic Markov chain, and the stationary solution in the 
thermodynamical limit is given by the condition:
\begin{equation}
  \left( X_{ln}-f_{ln} \right)^2 + p \left( X_{ln}-f_{1n}+f_{1,l-1} \right)^2 
    =   \label{cuarta}
\end{equation}
\[
   f_{ln} \left[ (1-p)(1-2X_{ln})+p(X_{1,l-1}-f_{1,l-1}-f_{1n})+f_{ln} \right] 
\]
for all $l>1$. For $l=1$ this condition reduces to
\begin{equation}
  (1+p)\left( X_{1n}-f_{1n} \right)^2 = (1-p)f_{1n}\left[1-2X_{1n}+f_{1n} 
     \right] \ .  \label{quinta}
\end{equation} 
Using Eq. ($\ref{cuarta}$) and Eq. ($\ref{quinta}$) we can calculate 
$f_{ln}$. The solutions reads
\[
   f_{ln} = \displaystyle{\frac{X^2_{ln} + \left( B_{n}-B_{n-1} \right)^2}
            {(1-p)/2 + p X_{ln} + B_{n} + B_{n-1}}} \ \ \ \ \ \ \ l \le n 
\]
with
\[
    B_n = \sqrt{ \left[ p X_{1n} + \frac{1-p}{4} \right]^2 - p X^2_{1n} 
           \left( \frac{1+p}{2}\right)} \ \ \ .
\]
Applying the definition, Eq.($\ref{tercera}$), we obtain the correlations:
\begin{equation}
 {\cal E}_{ij} = f_{ij}-f_{i,j-1}-f_{i+1,j}+f_{i+1,j-1}.   \label{sexta}
\end{equation}

\begin{figure}[tbp]
  \centerline{\hspace{-0.1cm}\psfig{figure=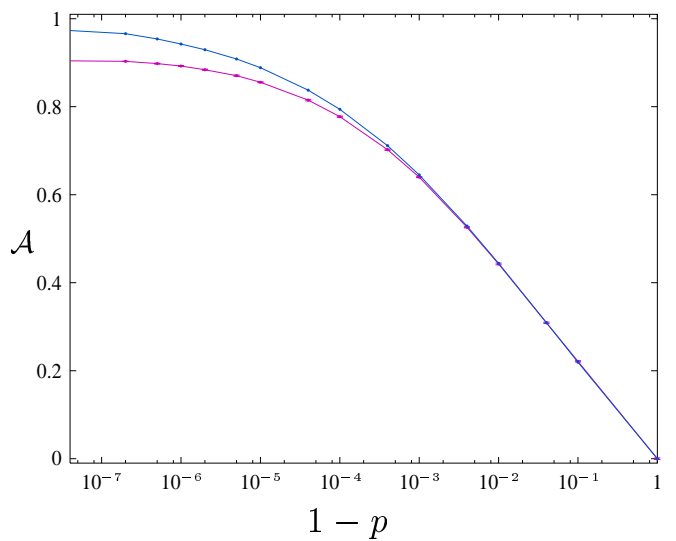,width=3.6in}}
    \caption{The lower curve corresponds to the measured assortativity 
             ${\cal A}$  of our simulations, whereas the upper curve 
             corresponds to the theory. We note that both curves 
             coincide for ${\cal A}<0.7$. Above this value the finite-size 
             corrections get important, leading to the measured value of 
             ${\cal A}<1$ for $p \to 1$.}
    \label{fig1}
\end{figure}

Finally, note that Eq.($\ref{sexta}$) reduces to the corresponding 
uncorrelated case ${\cal E}^r_{ij}$ when $p=0$, and reduces to
\begin{equation}
 {\cal E}_{ij} = \delta_{ij} \frac{iP(i)}{\langle i \rangle} 
\end{equation}
for the case $p=1$.


\section*{Simulation results}

In our simulations we apply our algorithm to the Barab\'asi-Albert network 
\cite{BA-model} with $N=10^5$ nodes and $L=2 \cdot 10^5$ links. We measure 
${\cal E}_{ij}$ as functions of $p$, and use them to calculate the 
corresponding values of ${\cal A}$. All simulation results are averaged over 
ten independent realizations of the algorithm as applied to the original 
network. 

Fig. \ref{fig1} shows the relation between the parameter $p$ and the 
coefficient of assortativity ${\cal A}$. The lower curve corresponds to the 
measured assortativity, and the upper to our theoretical prediction. Both 
curves coincide for ${\cal A}<0.7$. However, whereas the theoretical curve 
reach the value $1$ for $p\to 1$, the measured assortativity increases until 
the maximal value  smaller than one (${\cal A} \to 0.913$) is reached. This 
was theoretically expected, and is due to the finite-size effects mentioned 
above. 

\begin{figure}[tbp]
  \centerline{\hspace{-0.1cm}\psfig{figure=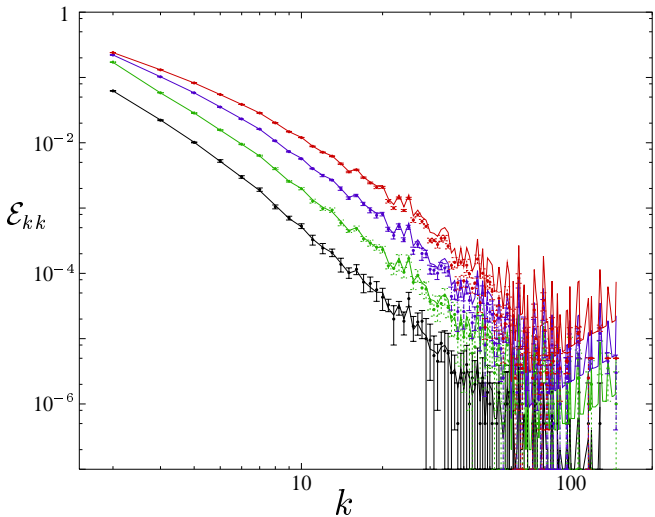,width=3.6in}}
    \caption{ ${\cal E}_{kk}$ as a function of $k$ for different values of 
             ${\cal A}$. From bottom to top:
             ${\cal A}= 0$, $0.221$, $0.443$, and $0.640$. The points are the 
             results of the simulations and the curves correspond to the 
             theory.}
    \label{fig2}
\end{figure}

To assess the goodness of the Eq. ($\ref{sexta}$) we compare in the 
Fig. \ref{fig2} the theoretical values of ${\cal E}_{kk}$, given by Eq. 
($\ref{sexta}$), with the simulations. The points correspond to the 
simulations and the curves are the corresponding theoretical results obtained
based on the actual degree distribution of a particular realization of the 
network discussed. We note that the agreement is really excellent.

\begin{figure}[tbp]
  \centerline{\hspace{-0.1cm}\psfig{figure=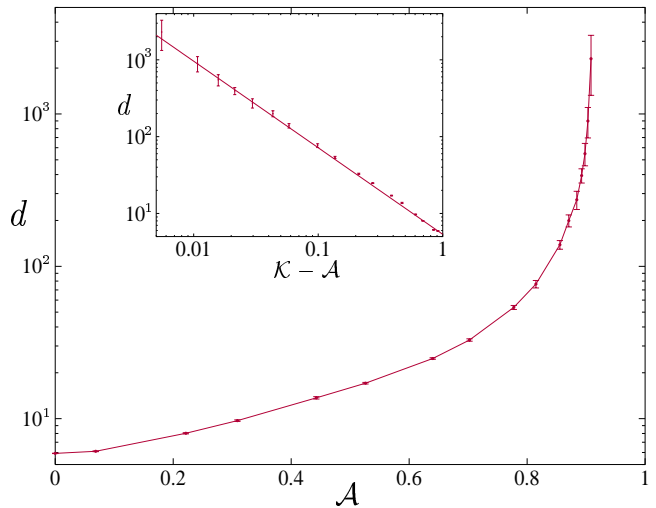,width=3.6in}}
    \caption{Diameter of the network versus coefficient of assortativity. We
             note that the diameter grows rapidly when ${\cal A}$ increases.
             In the inset the diameter is plotted on double logarithmic scales 
             as function of 
             ${\cal K} - {\cal A}$, being ${\cal K} = 0.931$. The slope of the
             straight line is 1.12.}
    \label{fig3}
\end{figure}

\emph{Diameter.---}
The diameter of a network is the average distance between every pair of 
vertices of the network, being defined as the number of edges along the 
shortest path connecting them. Uncorrelated scale-free networks 
show a very small diameter, typically growing as the logarithm of the network's
size. For networks with $N \simeq 10^5$ the diameter is about $d \simeq 6$. 
The results of the simulations show that the diameter grows rapidly when the 
assortativity of the network increases (fig. \ref{fig3}), so that it becomes 
hundred times larger than for the uncorrelated network when the coefficient of
assortativity tend to its maximal value. In the inset we plot the diameter as 
a function of ${\cal K}-{\cal A}$, where ${\cal K}=0.913$ corresponds to this 
maximal value of ${\cal A}$ attainable in the network. For our particular 
Barab\'asi-Albert network we thus have $d \propto (0.913-{\cal A})^{-1.12}$.

\emph{Clustering coefficient.---}
Clustering coefficients of a network are a measure of the number of loops 
(closed paths) of length three. The notion has its roots in sociology, where 
it was important to analyze the groups of acquaintances in which every 
member knows every other one. To discuss the concept of clustering, let us 
focus first on a vertex, having $k$ edges connected to $k$ other nodes termed 
as nearest neighbors. If these nearest neighbors of the selected node were 
forming a fully connected cluster of vertices, there would be $k(k-1)/2$ edges
between them. The ratio between the number of edges that really exist between 
these $k$ vertices and the maximal number $k(k-1)/2$ gives the value of the 
clustering coefficient of the selected node. The clustering coefficient of the
whole network $C$ is then defined as the average of the clustering 
coefficients of all vertices. One can also speak about the clustering 
coefficient of nodes with a given degree $k$, referring to the average of the 
clustering coefficients only over this type of nodes. We shall denote this 
degree-dependent clustering coefficient by $\bar{C}(k)$, to distinguish it 
from $C$. 
\begin{figure}[tbp]
  \centerline{\hspace{-0.1cm}\psfig{figure=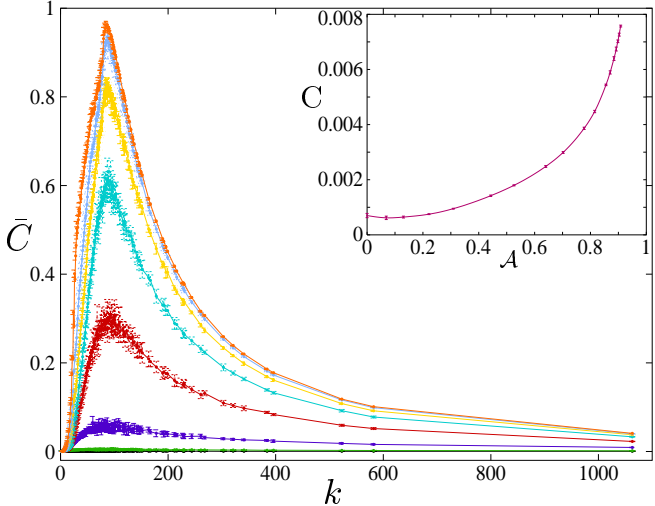,width=3.6in}}
    \caption{$\bar{C}(k)$ as a function of the degree of nodes $k$. Different
             curves correspond to different values of ${\cal A}$. From
             bottom to top ${\cal A}=0$, $0.069$, $0.221$, $0.443$, $0.640$,  
             $0.777$, $0.856$, and $1$. Inset: clustering coefficient $C$ 
             versus the degree of assortativity ${\cal A}$.}
    \label{fig4}
\end{figure}
Fig. \ref{fig4} shows the variation of both clustering coefficients with the 
assortativity of the network. The clustering coefficient $C$ increases with 
the assortativity (inset of the figure). The variation of $\bar{C}(k)$ shows 
more interesting features. The simulations show a peak around $k=90$ (probably
a finite size effect) whose height increases with the assortativity of the 
network. In the uncorrelated case $\bar{C}(k)$ does not depend on $k$ 
\cite{Clus}, but a strong tendency to clustering (for relatively large $k$) 
emerges when ${\cal A}$ grows. We also observe in our simulations that 
$\bar{C}(k=2)=0$ when ${\cal A} \simeq 1$ ($k=2$ correspond to the minimal 
degree of our vertices). This is not surprising since in a strongly 
assortative case almost all nodes with degree $k=2$ are connected between 
themselves, forming one or several large loops of length larger than three. 
This means that all nodes having this minimal degree (in our simulations the 
half of the total number of vertices) do not tend to contribute to the 
clustering coefficient $C$. 

In the present contribution we concentrate on the investigation of the  
properties of the proposed algorithm. However, we suggest, in relation 
to real networks, a simple modification of the algorithm, that perhaps could 
be useful. Thus, in order to generate assortativity only among highly 
connected vertices, one can apply the algorithm above only when at least one 
of the four nodes selected at the corresponding step has a degree larger than 
some chosen $k$. Provided all four nodes have a smaller degree, the the 
Maslov-Sneppen step is used. This procedure could lead to a larger value for 
the clustering coefficient, as it is observed in real networks ($C \ge 0.1$) 
\cite{baretalb}. The last ones might, however, have a much more intricate 
structure, partly governed by the metrics of the underlying space, as in the 
models discussed in \cite{Soki}, so that caution has to be exercised when 
applying results of theoretical models disregarding metrical relations to 
real networks.  
\begin{figure}[tbp]
  \centerline{\hspace{-0.1cm}\psfig{figure=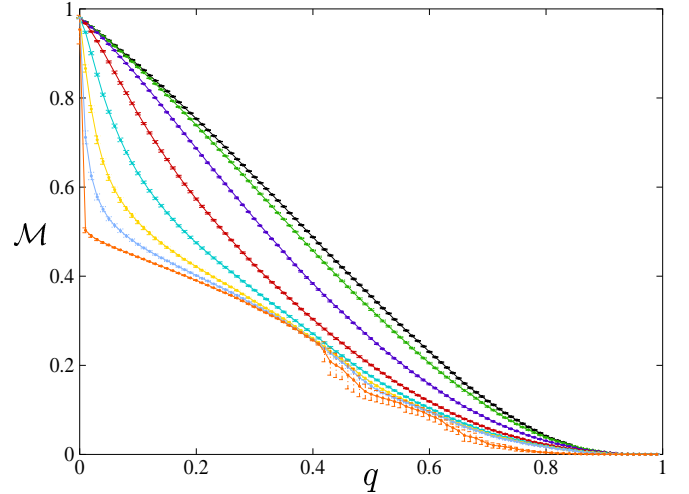,width=3.6in}}
    \caption{Fraction of nodes ${\cal M}$ in the giant component depending
             on the fraction of nodes removed from the network. The graph 
             compares the results for different degrees of assortativity. 
             From top to bottom: ${\cal A}=0$, $0.069$, $0.221$, $0.443$, 
             $0.640$, $0.777$, $0.856$, and $1$.}
    \label{fig5}
\end{figure}
\emph{Node percolation.---}
Node percolation corresponds to removal of a certain fraction of vertices from
the network, and is relevant when discussing their vulnerability to a random 
attack. Let $q$ be the fraction of the nodes removed. At a critical fraction
$q_c$, the giant component (largest connected cluster) breaks into tiny 
isolated clusters. Fig. \ref{fig5} shows the fraction of nodes ${\cal M}$ in
the giant component as a function of $q$ for different degrees of 
assortativity of the network. We note that the behavior of ${\cal M}(q)$ 
changes gradually with ${\cal A}$ from the uncorrelated case (upper curve) to
a quite different behavior when 
${\cal A} \rightarrow 1$ (lower curve), which indicates a very different 
topology in the network when it is strongly assortative. However, although the
particular form of the ${\cal M}$ dependence is different for different 
degrees of assortativity, the absence of the transition at finite 
concentrations ($q_c=1$) and the overall type of the critical behavior for 
correlated networks with the same  $P(k)$ seems to be the same as for 
uncorrelated networks, namely the one discussed in Refs. \cite{cohetal,ben}. 
We also point out that in case ${\cal A} \simeq 1$, a finite network is no 
longer fully connected: part of the nodes does not belong to the giant 
component even for $q=0$. The results suggest that, 
in the thermodynamical limit, the giant cluster at $q=0$ contains around a 
half of all nodes, and that its density then decays smoothly with $q$.

\section*{Conclusions}

In summary, we present an algorithm to generate assortatively correlated 
networks. In the termodynamical limit we obtain a theoretical expression for 
the generated correlations, which only depends on the degree distribution of 
the network and on the turnable parameter $p$ of the algorithm. Finally, we 
show that assortative correlations have a drastic influence on the statistical
properties of networks, changing strikingly their diameter and clustering 
coefficient, as well as their percolation properties. 

We also indicate that with a minor change in our algorithm one can produces
dissortative mixing too. The only change would be the following: after ordering
the nodes with respect to their degree, one rewires, with probability $p$,
the edges so that one link connects the highest connected node with the node 
with the lowest degree and the other link connects the two remaining vertices;
with probability $1-p$ one rewires the links randomly.

\section*{Acknowledgments}

Useful discussions with professor S. Havlin are gratefully acknowledged. IMS 
uses the possibility to thank the Fonds der Chemischen Industrie for the 
partial financial support.

\end{multicols}


\begin{references}

\bibitem{baretalb}  R. Albert and A.-L. Barab\'{a}si, Rev. Mod. Phys. 
   \textbf{74}, 47 (2002).
\bibitem{dorogov}  S.N. Dorogovtsev and J.F.F. Mendes, Adv. Phys. \textbf{51},
   1079 (2002).

\bibitem{NE}  M. E. J. Newman, Phys. Rev. E {\bf 67}, 026126 (2003).
\bibitem{Newm}  M. E. J. Newman, Phys. Rev. Lett. {\bf 89}, 208701 (2002).
\bibitem{Alex}  A. V\'azquez, M. Bogu\~n\'a, Y. Moreno, 
   R. Pastor-Satorras, and A. Vespignani, Phys. Rev. E {\bf 67}, 046111 (2003).
\bibitem{Cap}  A. Capocci, G. Caldarelli, and P. De Los Rios, Phys. Rev. E 
   {\bf 68}, 047101 (2003).
\bibitem{Pastor}  R. Pastor-Satorras, A. V\'azquez, and A. Vespignani,
   Phys. Rev. Lett. {\bf 87}, 258701 (2001).
\bibitem{Park}  M. E. J. Newman and J. Park, Phys. Rev. E {\bf 68}, 036112 
   (2003).
\bibitem{Berg}  J. Berg, and M. L\"assig, Phys. Rev. Lett. {\bf 89}, 228701
   (2002).
\bibitem{Goh}  K.-I. Goh, E. Oh, B.Kahng, and D. Kim, Phys. Rev. E {\bf 67},
   017101 (2003). 
\bibitem{Mas}  S. Maslov, and K. Sneppen, Science \textbf{296} 910 (2002).
\bibitem{Kra}  P. L. Krapivsky, and S. Redner, Phys. Rev. E {\bf 63} 
   066123 (2001).
\bibitem{Clus}  S. N. Dorogovtsev, Phys. Rev. E {\bf 69}, 027104 (2004).
\bibitem{Call}  D. S. Callaway, J. E. Hopcroft, J. M. Kleinberg, M. E. J. 
   Newman, and S. H. Strogatz, Phys. Rev. E {\bf 64}, 041902 (2001).
\bibitem{Juyong}  J. Park and M. E. J. Newman, Phys. Rev. E {\bf 68}, 026112 
   (2003).

\bibitem{Bog}  M. Bogu\~n\'a, R. Pastor-Satorras, and A. Vespignani, Phys.
   Rev. Lett. {\bf 90}, 028701 (2003).
\bibitem{Egui}  V. M. Egu{\'\i}luz, and K. Klemm, Phys. Rev. Lett. {\bf 89},
   108701 (2002).
\bibitem{Bo}  M. Bogu\~n\'a, and R. Pastor-Satorras, Phys. Rev. E {\bf 66},
   047104 (2002).
\bibitem{Schwartz}  N. Schwartz, R. Cohen, D. ben-Avraham, A.-L. Barab\'asi,
   and S. Havlin, Phys. Rev. E {\bf 66}, 015104 (2002).
\bibitem{Vaz}  A. V\'azquez, and Y. Moreno, Phys. Rev. E {\bf 67}, 015101
   (2003).
\bibitem{Mor}  Y. Moreno, J. B. G\'omez, and A. F. Pacheco, Phys. Rev. E 
   {\bf 68}, 035103 (2003).
 
\bibitem{Snep}  S. Maslov, K. Sneppen, and A. Zaliznyak, Physica A {\bf 333}, 
   529 (2004).

\bibitem{Rame}  A. Ramezanpour, V. Karimipour, and A. Mashaghi, Phys. Rev. E
   {\bf 67}, 046107 (2003).
\bibitem{Jo}  R. Xulvi-Brunet, W. Pietsch, and I. M. Sokolov, Phys. Rev. E 
   {\bf 68}, 036119 (2003).
\bibitem{Mar}  M. Bogu\~n\'a and R. Pastor-Satorras, Phys. Rev. E {\bf 68},
   036112 (2003).

\bibitem{BA-model}  A.-L. Barab\'{a}si, and R. Albert, Science \textbf{286},
   509 (1999).

\bibitem{Soki}  L.M. Sander, C.P. Warren, and I.M. Sokolov, Phys. Rev. E 
\textbf{66}, 056105 (2002).

\bibitem{cohetal}  R. Cohen, K. Erez, D. ben-Avraham, and S. Havlin, Phys.
Rev. Lett. \textbf{85}, 4626 (2000).
\bibitem{ben}  R. Cohen, D. ben-Avraham, and S. Havlin, Phys. Rev. E 
   \textbf{66}, 036113 (2002).

\end{references}
\end{document}